\documentclass{PoS}

\pdfoutput=1

\usepackage{amsmath}

\newcommand{\infinity}{\infty}
\newcommand{\be}{\begin{equation}}
\newcommand{\ee}{\end{equation}}
\newcommand{\bfig}{\begin{figure}}
\newcommand{\efig}{\end{figure}}
\newcommand{\bea}{\begin{eqnarray}}
\newcommand{\eea}{\end{eqnarray}}

\newcommand{\eq}[1]{Eq.~(\ref{#1})}
\newcommand{\pt}{$p_T$ }

\newcommand{\eqnpt}{p_T}

\newcommand{\lowpt}{low-\pt}

\newcommand{\highpt}{high-\pt}

\title{Testing AdS/CFT at LHC}

\ShortTitle{Testing AdS/CFT at LHC}

\author{\speaker{W.\ A.\ Horowitz}\\
        The Ohio State University, 191 W.\ Woodruff Ave., Columbus OH 43210, USA\\
        E-mail: \email{horowitz@mps.ohio-state.edu}}

\abstract{
After an introduction to jet phenomenology and tests of AdS/CFT at LHC we derive the heavy quark drag of a string dangling in a shock metric of AdS space, thus generalizing the AdS/CFT drag calculations in strongly coupled thermal media to momentum loss in both hot and cold nuclear matter.
}

\FullConference{High-pT Physics at LHC -09\\
		 February 4- 4 2009\\
		 Prague, Czech Republic}

\begin{document}

\section{Introduction}
The LHC will be a heavy ion physics discovery machine.  In particular the momentum regime opened up will usher in a new age of jet physics, which has the unique ability to probe the \highpt physics, \lowpt physics, and their mutual interaction in a bulk QGP medium.  And let there be no doubt that unless jets completely disappear due to initial state effects \cite{Tannenbaum:2008my}, not impossible given the recent PHENIX \highpt direct photon results \cite{Tannenbaum:2008my}, we will be able to exploit the observed data to learn about QCD: it is useful to remember that the ideas of fragility and surface emission have been experimentally and theoretically debunked (see, e.g., \cite{Horowitz:2006ya} and references therein).  

Fragility is loosely defined as the inability to invert experimental data using theoretical models to learn about the QGP medium properties.  Some early attempts at determining $\hat{q}$ found a large range of values were consistent with data.  Surface emission is loosely defined as the idea that the QGP is so opaque that the only jets that escape are produced at the outermost edge of the medium and suffer no energy loss.  Naively one argues that if surface emission exists then fragility necessarily follows; this is not quite true as, for radiative processes, the probability of no energy loss goes as $\exp(-N_g)$, where the $N_g$ average number of emitted gluons depends on the medium characteristics.  Nevertheless theoretical models using more sophisticated treatments of the energy loss and nuclear geometry do not exhibit surface emission; rather jets are seen to emerge from deep within the medium.  More damning are the recent statistical analyses by PHENIX which demonstrate a decided lack of fragility: current experimental error constrains the input parameter for the various theoretical models to within about 20\% \cite{Adare:2008cg}.  By the above argument, then, no fragility implies no surface emission.  

Part of the confusion with the issue of model sensitivity may possibly stem from plotting $R_{AA}$ as a function of the input parameter on a linear-linear scale.  It turns out that, to a good approximation, $R_{AA}$ has a power law dependence (e.g., for the ASW model doubling $\hat{q}$ approximately halves $R_{AA}$) \cite{Adare:2008cg}.  On a linear-linear scale this power-law shape looks misleadingly like an asymptotic approach to a nonzero value, that it exhibits fragility.  However plotting on a log-log scale makes its power law dependence manifest in a linear graph.

The above arguments against fragility and surface emission used results from models based on perturbative QCD (pQCD) energy loss.  And pQCD does a good job of explaining the normalization and momentum dependence of the single particle pion quenching as measured by $R_{AA}(\eqnpt)$.  However the consistency of ideal and nearly ideal hydrodynamics results with \lowpt data imply the QGP medium is strongly, rather than weakly coupled.  Moreover no two of the four separate measurements of $R_{AA}(\eqnpt)$ and $v_2(\eqnpt)$ for pions (the decay products of light quarks and gluons) and nonphotonic electrons (the decay products of heavy quarks) can be simultaneously quantitatively described by one set of input parameters in a perturbative energy loss model.  

On the other hand there are numerous instances of qualitative agreement between data and results based on the application of strong-coupling AdS/CFT techniques.  Before describing these it is worth noting that in many ways at temperatures not too far above $T_c$ QCD does not seem too different from the $\mathcal{N}=4$ super Yang-Mills (SYM) theory used in AdS/CFT derivations, and the controlling coupling parameter $\lambda = g_{YM}^2 N_c \sim 10 \gg 1$ is large.  With the known differences between the ``spherical cow'' approximation of QCD and actual QCD in mind, AdS/CFT has successfully described: the $\sim 3/4$ discrepancy between the entropy density slightly above $T_c$ and the Stefan-Boltzmann limit; the apparent small $\mathcal{O}(1/10)$ value of the shear viscosity to entropy density ratio $\eta/s$; the surprising suppression of nonphotonic electrons; and the conical emission pattern observed in two- and three-particle correlations (see \cite{Horowitz:2009pw} and references therein).

The AdS/CFT correspondence has been applied to \highpt jets in a number of ways (see, e.g., \cite{Horowitz:2009pw} and references therein).  One of these approximates a heavy quark as a string dangling in the fifth dimension, $z$.  Previous calculations \cite{Herzog:2006gh
} took the metric to have a black hole; the location of $z_H$, its event horizon, is related to the temperature of the thermalized medium through which the heavy quark propagates.  In \cite{Horowitz:2008zz} the momentum dependence of the double ratio of charm to bottom $R_{AA}$ was proposed as a robust test of the dominant heavy quark energy loss mechanism at LHC: for pQCD the ratio quickly approaches a value of 1; for AdS/CFT the ratio is essentially flat, given approximately by the ratio of the charm to bottom quark masses. 

Since the AdS/CFT conjecture is unproven and no dual to QCD yet exists it is difficult to test and possibly falsify its application.  In order to draw strong conclusions from comparing results derived from AdS/CFT and experiment one needs to test the generality of its predictions.  This may be done by using different field theories and alternate geometries.  Here we will pursue the latter and in the process generalize the work done previously with black hole metrics.  Specifically we embed the hanging string in a metric that represents a shock front medium.  The drag force then depends on the physical characteristics of the shock; it turns out that one only need consider the typical momentum scale of its constituents, $\Lambda$.  For $\Lambda = \sqrt{\pi} T$ we find exactly the same momentum loss as for a thermalized medium; since the shock can represent media for arbitrary $\Lambda$ our results are valid for energy loss in both hot and cold nuclear matter.  
\section{Shock Metric}\label{sec:met}
We will consider the generalized ``shock'' metric
\begin{align}
\label{lcmetric}
ds^2 & \equiv G_{\mu\nu}dx^\mu dx^\nu = \frac{L^2}{z^2}\left[ -2dx^+dx^- + 2\mu z^4\theta(x^-)dx^{-2} +dx_\perp^2+dz^2 \right] \\
\label{metric}
& = \frac{L^2}{z^2}\left[ -\left(1-\mu z^4\theta(x^-)\right)dt^2 -2\mu z^4 \theta(x^-)dtdx + \left(1+\mu z^4\theta(x^-)\right)dx^{2}+dx_\perp^2+dz^2 \right],
\end{align}
where we have used the $x^-=(x-t)/\sqrt{2}$ normalization of lightcone coordinates and dropped the $d\Omega_5^2$ standard metric of the five-sphere in AdS$_5\times S^5$.  Previous work used $\mu\delta(x^-)$ as the coefficient of the $dx^{-2}$ term in \eq{lcmetric} to investigate deep inelastic scattering (DIS) and the low energy aspects of heavy ion collisions in the strongly coupled regime \cite{Albacete:2008ze}.  $\mu$ represents a slightly different quantity in those papers with units GeV$^3$, unlike GeV$^4$ here.  As noted in \cite{Janik:2005zt} this coefficient can be any function of $x^-$, and we take it $\propto\theta(x^-)$ to represent an incoming dense medium of nuclear matter colliding with a heavy quark in its rest frame.  Usually lightcone coordinates mix time and the beam direction.  This is the case when applying our model in $p+A$ collisions; for $A+A$ collisions $x$ corresponds to the direction of motion of the heavy quark in the lab frame, which is usually transverse to the beam.

We are interested in the motion of a heavy quark in the background specified by the metric of \eq{metric} (for more details on this derivation, see \cite{Horowitz:2009pw}).  The test string action is $S_{NG} = -T_0\int d\tau d\sigma\sqrt{-g}$, where $g_{ab} = G_{\mu\nu}\partial_a X^\mu\partial_b X^\nu$.  Varying the action yields the equations of motion, $\nabla_a P^a_\mu = 0$, where $P^a_\mu = \pi^a_\mu/\sqrt{-g} = -T_0 \, G_{\mu\nu} \, 
\partial^a X^\nu$, and where $\nabla_a$ is the covariant derivative with respect to the induced metric, $g_{ab}$.

$X^\mu(\sigma^a)$ maps into the spacetime coordinates; choosing the static gauge, $\sigma^a=(t,z)$.  Assuming an asymptotic static solution $X^\mu(\sigma^a) = \xi(z)$ in a metric for which the shock exists over all space and time, the induced metric becomes time-independent.  The equations of motion reduce to $\partial \xi'/(z^4 \sqrt{-g})/\partial z = 0$.  Solving these for $\xi'$, with integration constant $C$, yields $\xi'(z) = \pm C z^2 \surd((1-\mu z^4)/(1-C^2 z^4))$.  There are two cases to consider for a string hanging from $z=0$ to $z=\infinity$ (solutions that turn around are considered in \cite{Horowitz:2009pw}): $C=0$ and $C\ne0$.  For $C\ne0$, the constant of integration is fixed by considering the signs of the numerator and denominator inside the radical as a function of $z$: for small $z$ both are positive; for large $z$ both are negative.  To avoid imaginary solutions, the numerator and denominator must change signs at precisely the same value of $z$; thus $C=\sqrt{\mu}$.  This leads to
\be
\label{zcubedsoln}
\xi(z) = x_0\pm\frac{\sqrt{\mu}}{3}z^3.
\ee
It is interesting to note that the near-boundary expansion of the static quark solution for the black hole metric (with horizon at $z=z_h$) is $x(t,z)\approx x_0\pm z^3/(3z_h^2)+vt$.  For $C=0$, $\xi=x_0$: the string hangs straight down.  Plugging this back into the action yields $S = -T_0 \int dt \int_{z_M}^\infinity dz \surd(1-\mu z^4)/z^2$.  The IR part of the $z$ integration gives the action an infinite imaginary part, and we interpret this as an infinitely unstable state that would immediately decay into the physical trailing string solution.
\section{Momentum Loss}\label{sec:momloss}
The drag force on the heavy quark in the SYM theory corresponds to the
momentum flow from the direction of heavy quark propagation down the
string, i.e., \ $dp/dt = -\pi^1_x$.  $\pi^a_\mu$ are the canonical
momenta:
\begin{align}
\begin{array}{cc}
\left( \begin{array}{c}
\pi^0_t \\
\pi^0_x \\
\pi^0_z \end{array} \right) = \frac{T_0 L^4}{z^4 \sqrt{-g}}
\left( \begin{array}{c}
-1-x'^2+\mu z^4(1-\dot{x}) \\
\dot{x}-\mu z^4(1-\dot{x}) \\
-x'\big( \dot{x} - \mu z^4 (1 - \dot{x}) \big)
\end{array}
\right); &
\left( \begin{array}{c}
\pi^1_t \\
\pi^1_x \\
\pi^1_z \end{array} \right) = \frac{T_0 L^4}{z^4 \sqrt{-g}}
\left( \begin{array}{c}
\dot{x}x'\\
-x'\\
-1+\dot{x}^2+\mu z^4(1-\dot{x})^2\\
\end{array}
\right).
\end{array}
\end{align}
The ``momentum change'' of our heavy quark solution given by
\eq{zcubedsoln}, where momentum change is in quotation marks as the
quark is held static, is then $dp/dt = - \pi^1_x = 2\pi\surd(\lambda\mu)$.

These derivations were made in the quark's rest frame; the above momentum loss is not that measured in the lab frame, the rest frame of the shock.  While the shock of the metric formally propagates at the speed of light, we think of it as an approximation to a medium slightly off the lightcone.  Then its rest frame is well defined, and we can relate $\mu$ to its properties via the holographic renormalization proceedure.

Following \cite{Albacete:2008ze}, we assume the medium is made up 
of $N_c^2$ valence gluons of the $\mathcal{N}=4$ SYM fields.  If in the rest frame of the medium the particles 
are isotropically distributed with a typical momentum of order $\Lambda$---with 
associated inter-particle spacing of order $1/\Lambda$---then the 00 component of the 
stress-energy tensor in the rest frame of the shock is $\langle T'_{00} \rangle \, \propto \, N_c^2 \Lambda^4$, where primes denote quantities in the rest frame of the medium and
proportionality is up to a constant numerical factor depending on the precise nuclear medium we choose to model.  Changing into lightcone
coordinates and boosting into the rest frame of the heavy quark yields $\langle T_{--} \rangle = N_c^2 \Lambda^4 \gamma^2 = N_c^2 \Lambda^4 (p'/M)^2$,
where we assumed ultrarelativistic motion for the heavy quark in the
medium rest frame, $p' \simeq M\gamma$. Comparing this with the EM tensor we read off $\mu = \pi^2 \Lambda^4 (p'/M)^2$.  We now need to find the relation between $dp/dt$ and $dp'/dt'$.

To do so note that $dp/dt$ is the 3-vector
component of the force 4-vector in the quark rest frame, $f^x \equiv dp/d\tau = dp/dt$. 
One sees that $\pi^1_t = 0$ ($f^t = 0$), and the 4-force boosted into the shock rest frame is $f'^x = -\gamma f^x = -\gamma dp/dt$, where the negative sign comes from boosting into a frame moving in 
the opposite direction.  From the definition of the 
4-force we also know that in this frame $f'^x \equiv dp'/d\tau = \gamma dp'/dt'$.  Therefore $dp/dt = -dp'/dt'$.  This leads us to our main result:
\begin{align}
  \label{momgaintwo}
  \frac{d p'}{d t'} \, = \, - \frac{\sqrt{\lambda}}{2} \,
  \frac{\Lambda^2}{M_q} \, p'.
\end{align}
Should we take the typical momentum to be $\Lambda = \sqrt{\pi} T$ then our result exactly
reproduces that of the black hole metric, $d p' / d t'
= - \pi \sqrt{\lambda} \, T^2 \, p'/ (2 \, M_q)$
\cite{Herzog:2006gh
}.  This makes sense because if one infinitely boosts the standard black hole metric while keeping $\gamma^2/z_h^4$ fixed\footnote{We wish to thank Alberto Guijosa for pointing out this boosting procedure.} the shock metric, \eq{metric}, is recovered with $\mu = \pi^4 \gamma^2 T^4$.
\section{Speed Limit of Applicability}\label{sec:limits}
As shown in
\cite{Herzog:2006gh
}, there are
limits to the applicability of the heavy quark drag calculations in a
black hole metric.  Reality of the point particle action in 5 bulk dimensions
in the AdS BH metric with the horizon at $z=z_h$ requires that
$\sqrt{-G_{\mu\nu} \, (dx^\mu/d\tau) \,  (dx^\nu/d\tau)} =
\sqrt{(1-(z/z_h)^4)/z^2-v^2/z^2}$ be real.  Then $v^2<1-(z/z_h)^4$
leading to $\gamma< (z_h/z)^2 < (z_h/z_M)^2$.  While the metric, \eq{metric}, does not support an event horizon,
reality of the point particle action still yields an asymmetric,
$z$-dependent result for the local speed of light; to wit, in the rest
frame of the heavy quark $(\mu z^4-1)/(\mu z^4 + 1) \le v \le 1$.
At the boundary, $z=0$, the usual Minkowski speed of light limit,
$-1\le v\le 1$, is recovered. Motion at the stack of D3 color branes
(at $z=\infty$) is restricted to the speed of light in the direction
of the shock medium motion.  In a similar way one finds that reality
of the Nambu-Goto action requires the velocity of the string at
$z=\infinity$ to be $+1$.

The speed limit creates an argument against
the reversed trailing string solution (given by the RHS of \eq{zcubedsoln} taken with a minus sign). The only classical fluctuations with non-zero velocity supported by
the time-reversed static string solution are those that, at $z=\infinity$, give motion at the speed of light in the
direction of the shock motion, i.e., towards the physical trailing
string solution (given by \eq{zcubedsoln} with the plus sign on the RHS).

Plugging $v=0$ for our static quark into the above speed limit gives us the bound
for this calculation $\mu \, z_M^4 \le 1$, where $z_M = \sqrt{\lambda}/(2\pi M_q)$ is found by integrating out the $z$ coordinate in the string action and demanding its equality to the point particle action of a quark of mass $M_q$.  For $\mu=\pi^2\Lambda^4\gamma^2$, along with $\Lambda = \sqrt{\pi} \,
T$, we obtain
\begin{align}
  \gamma \le \frac{4 \, \pi \, M_q^2}{\lambda \, \Lambda^2} = \frac{4
    \, M_q^2}{\lambda \, T^2}
\end{align}
The speed limit in this geometry is therefore identical to
that for the BH metric \cite{Herzog:2006gh
}. 
\section{Conclusions}
We used the AdS/CFT correspondence to calculate the momentum loss of a heavy quark in a strongly coupled medium represented by a shock.  This drag depends only on a single scale parameter describing the medium, $\Lambda$.  Taking $\Lambda=\sqrt{\pi} T$ reproduces the previous work done in a black hole metric.  Since the scale can be arbitrary we have generalized these results to media of any isotropic distribution.  It turns out that, just as for perturbative energy loss, the form of the drag is independent of medium thermalization.  Unfortunately the momentum speed limit known from the thermal cases persists; it is precisely the same in the shock for $\Lambda = \sqrt{\pi} T$.  We argued that this investigation of an alternative geometry to the usual black hole metric gives us greater confidence in the robustness of the application of this AdS/CFT methodology to heavy ion physics, and hence also greater confidence in comparing---and possibly falsifying---the AdS/CFT predictions with data.  We note that a convincing comparison at LHC will crucially require a separate $p+Pb$ run or a direct photon measurement in $Pb+Pb$ to disentangle possible initial from final state effects.

\vspace{-.05in}

\section{Acknowledgments}
We are grateful to Alberto Guijosa for enlightening discussions.  This work is supported by the Office of Nuclear Physics, of the Office of Science, of the U.S.\ Department of Energy under Grant No.\ DE-FG02-05ER41377.

\end{document}